# Fabrication of uniformly doped graphene quantum Hall arrays with multiple quantized resistance outputs

S. M. Mhatre,[1,2] N. T. M. Tran,[1,3] H. M. Hill,[1] C.-C. Yeh,[1,4] D. Saha,[1] D. B. Newell,[1] A. R. Hight Walker,[1] C.-T. Liang,[2,4] R. E. Elmquist,[1] and A. F. Rigosi[1,a]

[1]*Physical Measurement Laboratory, National Institute of Standards and Technology (NIST), Gaithersburg, Maryland, 20899-8171, USA*

[2]*Graduate Institute of Applied Physics, National Taiwan University, Taipei, 10617, Taiwan*

[3]*Joint Quantum Institute, University of Maryland, College Park, MD 20742, USA*

[4]*Department of Physics, National Taiwan University, Taipei 10617, Taiwan*

In this work, limiting factors for developing metrologically useful arrays from epitaxial graphene on SiC are lifted with a combination of centimeter-scale, high-quality material growth and the implementation of superconducting contacts. Standard devices for metrology have been restricted to having a single quantized value output based on the $v$ = 2 Landau level. With the demonstrations herein of devices having multiple outputs of quantized values available simultaneously, these versatile devices can be used to disseminate the ohm globally. Such devices are designed to give access to quantized resistance values over the range of three orders of magnitude, starting as low as the standard value of approximately 12.9 kΩ and reaching as high as 1.29 MΩ. Several experimental methods are used to assess the quality and versatility of the devices, including standard lock-in techniques and Raman spectroscopy.

---

[a] Author to whom correspondence should be addressed.  Mail: Albert Rigosi, MS 8171, 100 Bureau Drive, NIST, Gaithersburg, MD 20899.

Because of graphene's unique properties [1-4], it can be utilized in a wide variety of academic and commercial pursuits. When grown on a 4H-SiC substrate, epitaxial graphene (EG) has been developed into devices and utilized for the advancement of resistance metrology because of the robust quantum Hall effect (QHE) exhibited across a highly applicable range of magnetic fields ($B$-fields). The main requirement for such devices to be successfully implemented as standards is that the exhibited resistance be well-quantized [5-10]. Most graphene-based standards that have recently been implemented operate at the resistance plateau formed by the $v = 2$ plateau ($\frac{1}{2}\frac{h}{e^2} = \frac{1}{2}R_K \approx 12906.4037$ Ω, where $h$ is the Planck constant and $e$ is the elementary charge), with other efforts using the $v = 6$ plateau [11].

In recent years, graphene-based standards have been fabricated in the hopes of expanding the available quantized resistances useable for metrological purposes. Within the International System of Units (SI), the unit of the ohm has been historically disseminated from a single value of the QHE ($v = 2$), and this constraint heavily restricts the infrastructure and equipment with which one may disseminate the recently redefined quantum SI. The effort for accessing new values is thus highly beneficial for electrical standards. Thus far, several demonstrations of multiple Hall bars in parallel, series, or arranged as $p$-$n$ junctions have been performed to create resistance values of $qR_K$ where $q$ is a positive rational number [12-21]. The limiting factors in how these standards are developed primarily include the area over which high-quality EG may be grown, as well as the device restriction of outputting one value, even if that value is one of a set of many possible values ($qR_K$) [22-23]. Thus, to further improve on how standards can be used to disseminate the ohm globally, demonstrations of devices that could output more than one value are highly desirable.

In this work, we demonstrate one such device design that confirms the feasibility of having multiple outputs of quantized values available simultaneously over the range of three orders of magnitude, starting as low as the standard value of approximately 12.9 kΩ and reaching as high as 1.29 MΩ, or the equivalent of 100 Hall bars in series. Several experimental methods are used to assess the quality and versatility of the devices, including standard lock-in techniques and Raman spectroscopy.

Wafers of 4H-SiC that have been chemically and mechanically smoothed along the Si-face were obtained from CREE and diced into 7.7 mm × 7.7 mm chips (see Acknowledgments). All chips were then cleaned with a Piranha solution (3:1 $H_2SO_4$:$H_2O_2$) for 2 h followed by a 5 min rinse with 51 % hydrofluoric acid (diluted with deionized water). Just prior to growth, the chips were coated with a dilute solution of carbon-based photoresist (AZ 5214E, see Acknowledgments) in isopropyl alcohol to utilize polymer-assisted sublimation growth (PASG) [24]. The Si-face side of each chip was pressed against a polished glassy carbon slab (SPI Glas 22, see Acknowledgments) such that interference rings were observed. This ensures that the surfaces are closely spaced (up to 2 μm) to limit Si atoms from escaping, thus improving graphene uniformity. The growth furnace was flushed with Ar gas and filled to about 103 kPa from a 99.999 % liquid argon source. The graphite-lined resistive-element furnace (Materials Research Furnaces Inc., see Acknowledgments) was held at 1850 °C for 3 min to 4 min. The furnace heating and cooling rates were about 1.5 °C/s. A defect-free hexagonal lattice of carbon atoms, EG, is formed on top of a covalently bonded carbon buffer layer when 4H-SiC substrates are heated to these temperatures [25-26].

The grown EG films were characterized by means of optical and confocal laser scanning microscopy (CLSM) to select those with monolayer coverage greater than 99 %, minimal bilayers, and uniform SiC step heights less than 1 nm. For device fabrication, the EG layer was protected by a 20 nm layer of Pd/Au, followed by a photolithography process that defines the Hall bar and device contact pattern [27-29]. For electrical contacting of the array devices, a 100 nm layer of superconducting NbTiN was deposited over the contacts to form device interconnects that greatly improve array performance, eliminated any undesired interconnection resistance, and allow for two-terminal measurements [13, 20]. The separation of the NbTiN layer and the EG was greater than 80 nm to prevent undesired quantum effects such as Andreev reflection. Some EG was grown without PASG pre-processing, resulting in larger parallel SiC steps (1 nm to 5 nm) and greater than 99 % monolayer graphene, enabling us to quantify the influence of the steps themselves.

Gateless control of the carrier density was accomplished by implementing a functionalization process using $Cr(CO)_3$ due to its successful use in other studies [30-32]. A small nitrogen-filled furnace at 130 °C was used to initiate the hexahapto functionalization [$(\eta^6$-graphene)-$Cr(CO)_3$] of EG. This process ensures a uniform carrier density across the surface of EG. The typical value for the carrier density after ambient atmospheric exposure is around $10^{10}$ cm$^{-2}$ [32], which can be compared to the typical values of inherent doping in EG of $10^{13}$ cm$^{-2}$ [33]. An illustration of the device design is shown in Fig. 1 (a) along with corresponding CLSM images of two distinct array elements in Fig. 1 (b) and (c). Each individual element in the array functions as a 12.9 kΩ standard Hall bar and has a multiple-branch design required to optimize the current flow and eliminate the effect of contact resistances [13, 20]. The contrast achieved with CLSM confirmed the high-quality of the EG. The mid-array device (51[st] element) is meant to serve as a diagnostic tool in the event of poor quantization. It may also be used with a high-resistance cryogenic current comparator that could enable calibrations of higher values of resistance. A large contact pad was fabricated at the end of every row (10 elements), enabling access to quantized values in the multiples of 129 kΩ, up to and including 1.29 MΩ. These values are summarized in Fig. 1 (a).

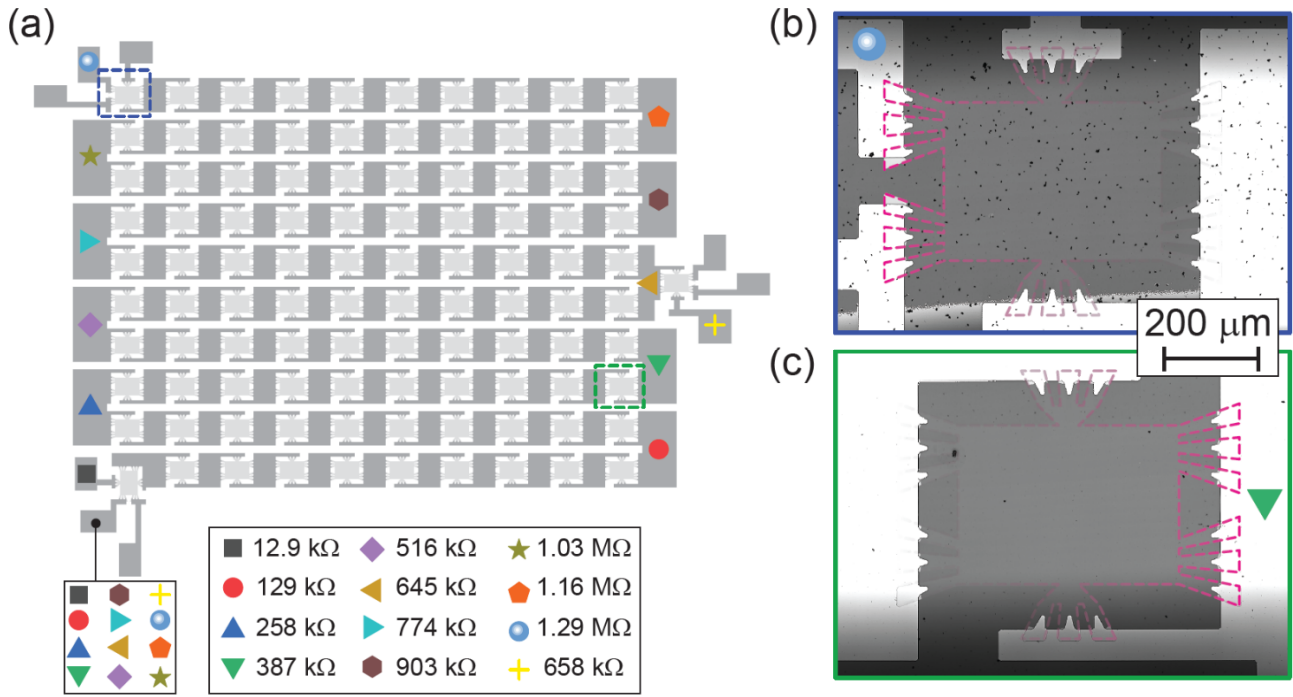

FIG. 1. (a) An illustration of the 1.29 MΩ Hall array device is shown with locations for measurements of subarrays that provide various multiples of quantized Hall resistance. Two example areas, outlined by the dashed blue and green lines, are inspected with confocal laser scanning microscopy. (b) and (c) The two inspections reveal the uniformity of the graphene growth over the near-centimeter-wide SiC substrate. A fading magenta outline is overlaid to show the boundary of the Hall bar elements. The scattering of debris varies across the device, but the debris does not interact with the graphene and is composed of remnants from the functionalization process.

Raman spectroscopy was also used to confirm the uniformity of the EG layer since visual inspections may not identify flaws in the physical properties of the grown EG film. Measurements were performed with a Renishaw InVia micro-Raman spectrometer (see Acknowledgements) using a 633 nm wavelength excitation laser source. The spectra were measured and collected using a backscattering configuration, 1 µm spot size, 300 s acquisition time, 1 mW power, 50 × objective, and 1200 mm$^{-1}$ grating. For statistics on the EG quality, rectangular Raman maps were collected with step sizes of 1 µm in a 25 by 25 raster-style grid and repeated on various elements of the array devices. A summary of the Raman analysis is provided in Fig. 2. Within the device, whose optical image is shown in Fig. 2 (a), two regions separated by a maximum distance are inspected closely. An example of a Raman spectrum of the 2D (G') mode is shown in Fig. 2 (b) and used as the primary metric for comparing EG quality across the regions. Note that the D and G peaks were not selected due to their obscurity caused by strong optical responses by the SiC substrate.

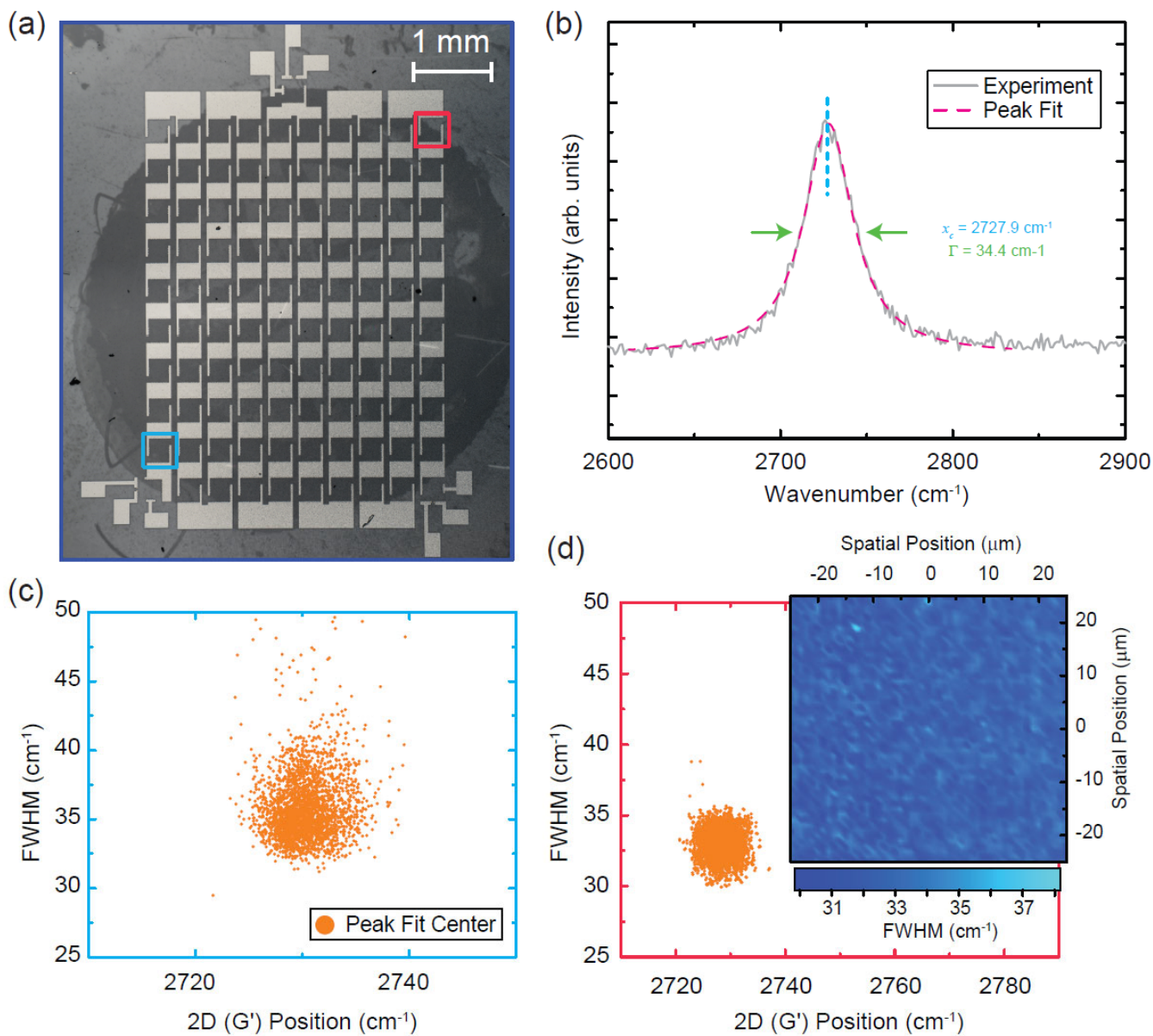

FIG. 2. (a) An optical image of the full device is shown with a cyan and red region indicating the two example elements whose Raman map results are shown in (c) and (d), respectively. (b) An example Raman spectrum focuses on the 2D (G') mode of graphene since the D and G modes have signals that are overwhelming by neighboring optical responses from the SiC substrate. Each measured peak is fitted with a Lorentzian profile so that the peak center and FWHM are extracted. (c) The array element outlined in cyan in (a) was inspected more closely. The scatterplot shows the distribution of FWHM with 2D mode peak centers. (d) A repeated analysis is performed for the array element outlined in red in (a). In addition to the scatterplot, a color map of the position-dependent FWHM is shown to verify the uniformity of the EG film.

The peak in Fig. 2 (b) and all additional 2D peaks were fitted with a Lorentzian profile (dashed magenta) to extract a peak position and full-width-at-half-maximum (FWHM). Every spectrum from each Raman map was analyzed in the same fashion, yielding the two scatterplots in Fig. 2 (c) and (d). The latter panel also contains a real space distribution of values for the FWHM to give an example of the variation expected within the region. These data verify the length scales on which EG can be grown with excellent quality. After an optical verification of film quality, the next necessary step for such devices is to assess their transport properties. Quantum Hall transport measurements were performed in a Janis Cryogenics $^4$He cryostat. All devices were mounted onto a transistor outline (TO-8) package, and all corresponding data were collected between magnetic field values of 0 T and ± 9 T to characterize the magnetoresistances of the devices. All measurements were performed at approximately 1.5 K with source-drain currents either as 100 nA or 500 nA. Prior to cooldown, devices were annealed in vacuum as described in Ref. [32] to obtain a desired electron density corresponding to a ν = 2 plateau onset of approximately 4 T.

An exemplary set of transport measurements is shown in Fig. 3 (a). Most allowable values were measured as labeled in Fig. 1, where the limit of what could be measured stems mostly from the contact pad's lateral size, which lends itself for wirebonding. Though many intermediate values could be bonded within the array, alternate designs are recommended since potential damage to one element along the array would likely split the array into two separate, neighboring devices. For the magnetoresistances in Fig. 3, all were collected with lock-in amplifiers. Though this technique provides the advantage of collecting data swiftly and with higher magnetic field resolution, it does introduce minor errors due to equipment impedances in the MΩ range. To account for this error, the values of each plateau were verified more precisely with a digital multimeter at sufficiently high magnetic field (greater than 5 T). Verification of these values to between parts in $10^3$ and parts in $10^6$ with a digital multimeter enabled for the correction of these data via vertical offset translation and are shown as a group in Fig. 3 (b).

The noise in each plateau from Fig. 3 is relatively symmetric for both magnetic field polarities and may be associated with the lock-in amplifier technique limitations, and, to some degree, a result of the low current injected into the device. Any noise coming from the latter was minimized by increasing the injected current, as seen in Fig. 3 (c), where the two currents of 100 nA and 500 nA are compared. The justification for comparing two different currents has more to do with the eventual compatibility of these devices with metrological infrastructure. This is also the reason that a 51$^{st}$ element was introduced in the array. For the scope of this work however, the 51$^{st}$ device simply serves as an additional support that the whole array device may be adequately quantized. It should be noted that although higher applied currents typically result in less noise and better compatibility with metrological equipment, there is a point at which excessive Joule heating at various hot spots within the quantum Hall array would cause the device to lose its adequate quantization, leading to an even larger error when measuring the plateau's values.

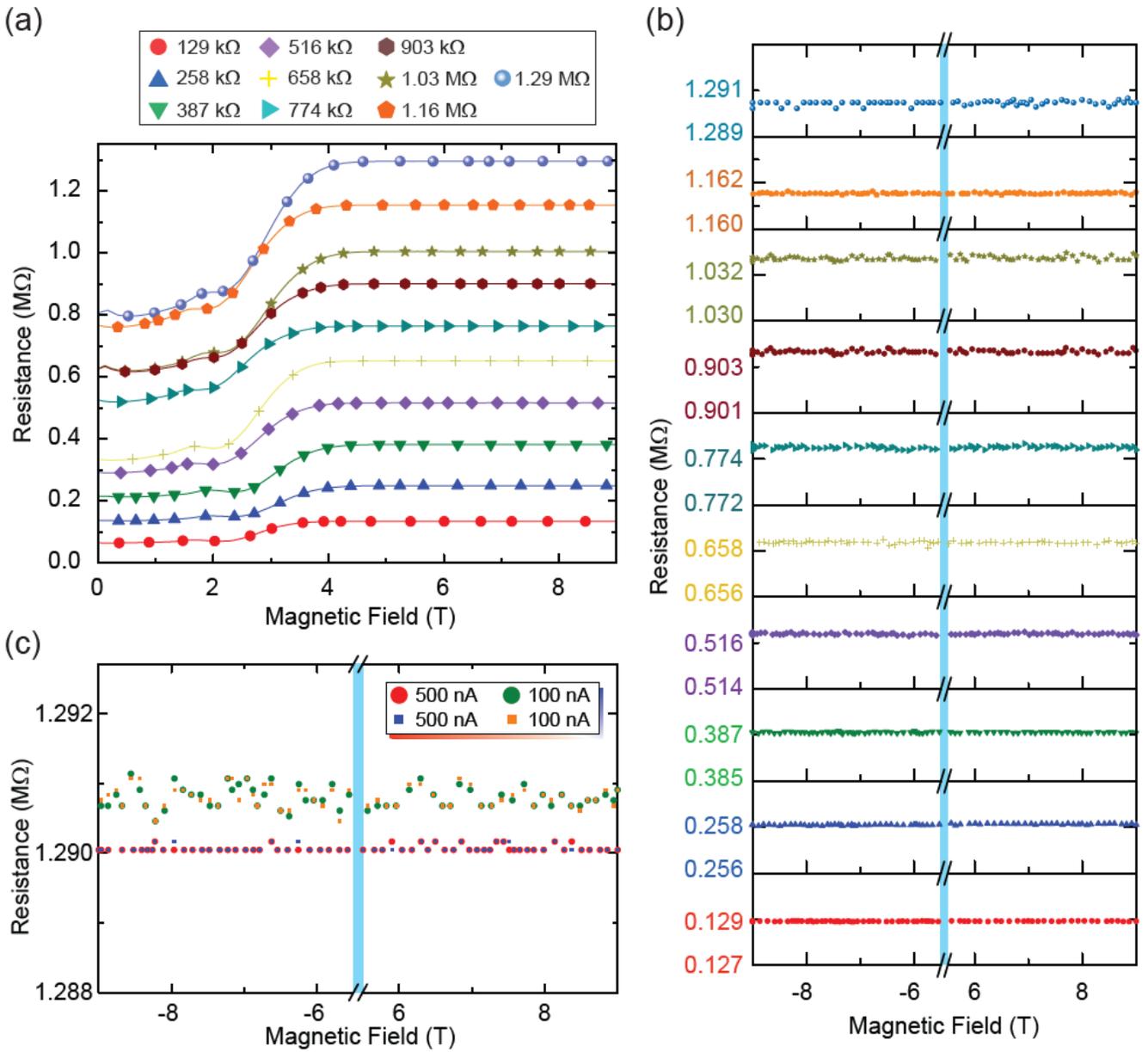

FIG. 3. (a) A full summary of the transport measurements for each of the designated labels from figure one is shown. The magnetoresistances were collected with lock-in amplifiers. Although this technique lends the advantage of collecting data swiftly and finely, it also potentially introduces minor errors due to equipment impedance. To account for this error, the values of each plateau were verified more precisely with a digital multimeter at sufficiently high magnetic field (greater than 5 T). Verification of these values to within a part in $10^6$ enabled for the correction of these data via offset translation. (b) Each plateau is shown magnified at both magnetic field polarities to demonstrate the level of noise introduced by the lock-in amplifier technique. Such noise was minimized by increasing the injected current, as seen in (c), where the two currents of 100 nA and 500 nA are compared.

The versatility of these devices has been demonstrated and our samples are encouraged to be tested for metrological purposes. Overall, in this work, the boundaries for limiting factors in developing quantum Hall arrays were removed to the point that several orders of magnitude of measurable quantized resistance could be measured. This advance was feasible due to the combination of centimeter-scale, high-quality material growth and the implementation of superconducting contacts.

Specifically, the standard Hall value at the ν = 2 plateau of approximately 12.9 kΩ, was used as a building block to reach values as high as 1.29 MΩ. Devices were inspected for superior material quality by means of Raman spectroscopy, optical microscopy, and confocal laser scanning microscopy. Applications to any device manufacturing requiring outputting of more than a single quantized value of resistance will benefit from this demonstration. It now stands that more complex arrays may be designed and fabricated, while keeping the several experimental methods described herein as ways to assess the quality of those devices.

## ACKNOWLEDGMENTS

SMM and AFR designed the experiment and, with NTMT and CCY, collected transport data. HMH and ARHW provided assistance with Raman spectroscopy. SMM, NTMT, and DS fabricated devices. DBN, CTL, and REE provided general project oversight and guidance. The manuscript was written with contributions from all authors. The authors thank T. Mai, G. Fitzpatrick, A. L. Levy, and E. C. Benck, for assistance with the internal NIST review process. The authors declare no competing interest.

Commercial equipment, instruments, and materials are identified in this paper in order to specify the experimental procedure adequately. Such identification is not intended to imply recommendation or endorsement by the National Institute of Standards and Technology or the United States government, nor is it intended to imply that the materials or equipment identified are necessarily the best available for the purpose

## REFERENCES

[1] A.K. Geim and K.S. Novoselov, Nat. Mater. **6**, 183 (2007).

[2] K.S. Novoselov, V.I. Fal'ko, L. Colombo, P.R. Gellert, M.G. Schwab, and K. A. Kim, Nature **490**, 192 (2012).

[3] A.H. Castro Neto, F. Guinea, N. M. R. Peres, K. S. Novoselov, and A. K. Geim, Rev. Mod. Phys. **81**, 109 (2009).

[4] S. Das Sarma, S. Adam, E.H. Hwang, and E. Rossi, Rev. Mod. Phys. **83**, 407 (2011).

[5] T. Oe, A. F. Rigosi, M. Kruskopf, B. Y. Wu, H. Y. Lee, Y. Yang, R. E. Elmquist, N. H. Kaneko, D. G. Jarrett, IEEE Trans. Instrum. Meas. **69**, 3103-8 (2019).

[6]R. Ribeiro-Palau, F. Lafont, J. Brun-Picard, D. Kazazis, A. Michon, F. Cheynis, O. Couturaud, C. Consejo, B. Jouault, W. Poirier, and F. Schopfer, Nature Nanotechnol. **10**, 965 (2015).

[7]A. Tzalenchuk, S. Lara-Avila, A. Kalaboukhov, S. Paolillo, M. Syväjärvi, R. Yakimova, O. Kazakova, T.J.B.M. Janssen, V. Fal'ko, and S. Kubatkin, Nat Nanotechnol. **5**, 186 (2010).

[8]A.F. Rigosi and R.E. Elmquist, Semicond. Sci. Technol. **34**, 093004 (2019).

[9]B. Jeckelmann, B. Jeanneret. Rep. Prog. Phys. **64**, 1603 (2001).

[10]H. He, S. Lara-Avila, T. Bergsten, G. Eklund, K. H. Kim, R. Yakimova, Y. W. Park, S. Kubatkin. Conference on Precision Electromagnetic Measurements (CPEM 2018) 1-2 (2018).

[11]A. R. Panna, I-F. Hu, M. Kruskopf, D. K. Patel, D. G. Jarrett, C.-I Liu, S. U. Payagala, D. Saha, A. F. Rigosi, D. B. Newell, C.-T. Liang, R. E. Elmquist, Phys. Rev. B **103**, 075408 (2021).

[12]S. Novikov, N. Lebedeva, J. Hamalainen, I. Iisakka, P. Immonen, A. J. Manninen, and A. Satrapinski, J. Appl. Phys. **119**, 174504 (2016).

[13]M. Kruskopf, A. F. Rigosi, A. R. Panna, D. K. Patel, H. Jin, M. Marzano, M. Berilla, D. B. Newell, and R. E. Elmquist, IEEE Trans. Electron Dev. **66**, 3973 (2019).

[14]J. Hu, A. F. Rigosi, J. U. Lee, H.-Y. Lee, Y. Yang, C.-I Liu, R. E. Elmquist, and D. B. Newell, Phys. Rev. B **98**, 045412 (2018).

[15]F. Delahaye, J. Appl. Phys. **73**, 7914 (1993).

[16]A. F. Rigosi, D. Patel, M. Marzano, M. Kruskopf, H. M. Hill, H. Jin, J. Hu, A. R. Hight Walker, M. Ortolano, L. Callegaro, C.-T. Liang, and D. B. Newell, Carbon **154**, 230 (2019).

[17]A. Lartsev, S. Lara-Avila, A. Danilov, S. Kubatkin, A. Tzalenchuk, and R. Yakimova, J. Appl. Phys. **118**, 044506 (2015).

[18]J. Hu, A. F. Rigosi, M. Kruskopf, Y. Yang, B. Y. Wu, J. Tian, *et al.* Sci. Rep., **8**, 15018 (2018).

[19]J. Park, W. S. Kim, D. H. Chae, Appl. Phys. Lett. **116**, 093102 (2020).

[20]M. Kruskopf, A. F. Rigosi, A. R. Panna, M. Marzano, D. Patel, H. Jin, D. B. Newell, and R. E. Elmquist, Metrologia **56**, 065002 (2019).

[21]Z. S. Momtaz, S. Heun, G. Biasiol, S. Roddaro, Phys. Rev. Appl. **14**, 024059 (2020).

[22]M. Woszczyna, M. Friedmann, T. Dziomba, T. Weimann, and F.J. Ahlers, Appl. Phys. Lett. **99**, 022112 (2011).

[23]A.F. Rigosi, D.K. Patel, M. Marzano, M. Kruskopf, H.M. Hill, H. Jin, J. Hu, R. E. Elmquist, and D.B. Newell, Physica B **582**, 411971 (2019).

[24]M. Kruskopf, D. M. Pakdehi, K. Pierz, S. Wundrack, R. Stosch, T. Dziomba., M. Götz, J. Baringhaus, J. Aprojanz, and C. Tegenkamp, 2D Mater. **3**, 041002 (2016).

[25]T. Seyller, A. Bostwick, K. V. Emtsev, K. Horn, L. Ley, J. L. McChesney, T. Ohta, J. D. Riley, E. Rotenberg, F. Speck, Phys. Status Solidi (b) **245**, 1436-46 (2008).

[26]H. M. Hill, A. F. Rigosi, S. Chowdhury, Y. Yang, N. V. Nguyen, F. Tavazza, R. E. Elmquist, D. B. Newell, and A. R. Hight Walker, Phys. Rev. B **96**, 195437 (2017).

[27]A. F. Rigosi, H. M. Hill, N. R. Glavin, S. J. Pookpanratana, Y. Yang, A. G. Boosalis, J. Hu, A. Rice, A. A. Allerman, N. V. Nguyen, C. A. Hacker, R. E. Elmquist, A. R. Hight Walker, and D. B. Newell, 2D Mater. **5**, 011011 (2017).

[28]A. F. Rigosi, C.-I. Liu, B.-Y. Wu, H.-Y. Lee, M. Kruskopf, Y. Yang, H. M. Hill, J. Hu, E. G. Bittle, J. Obrzut, A. R. Hight Walker, R. E. Elmquist, and D. B. Newell, Microelectron. Eng. **194**, 51-5 (2018).

[29]A. F. Rigosi, N. R. Glavin, C.-I. Liu Y. Yang, J. Obrzut, H. M. Hill, J. Hu, H.-Y. Lee, A. R. Hight Walker, C. A. Richter, R. E. Elmquist, and D. B. Newell, Small **13**, 1700452 (2017).

[30]E. Bekyarova, S. Sarkar, S. Niyogi, M.E. Itkis, and R. C. Haddon, J. Phys. D: Appl. Phys. **45**, 154009 (2012).

[31]S. Che, K. Jasuja, S. K. Behura, P. Nguyen, T. S. Sreeprasad, and V. Berry, Nano Lett. **17**, 4381 (2017).

[32]A. F. Rigosi, M. Kruskopf, H. M. Hill, H. Jin, B.-Y. Wu, P. E. Johnson, S. Zhang, M. Berilla, A. R. Hight Walker, C. A. Hacker, D. B. Newell, and R. E. Elmquist, Carbon **142**, 468 (2019).

[33]T. J. B. M. Janssen, A. Tzalenchuk, R. Yakimova, S. Kubatkin, S. Lara-Avila, S. Kopylov, et al. Phys. Rev. B **83**, 233402 (2011).